\documentclass{ws-procs9x6}

\begin{document}
\title{Constraint on $\Delta g(x)$ at large $x$}

\author{M. Hirai, S. Kumano$^{*}$, and N. Saito$^\dagger$\\
(Asymmetry Analysis Collaboration)}
\address{
Department of Physics,         
Tokyo Institute of Technology \\
2-12-1, Ookayama, Meguro-ku,Tokyo,     
152-8550, Japan                \\
E-mail: mhirai@th.phsy.titech.ac.jp}

\address{$^*$
Institute of Particle and Nuclear Studies \\
High Energy Accelerator Research Organization (KEK) \\
1-1, Ooho, Tsukuba, Ibaraki, 305-0801, Japan \\
and Department of Particle and Nuclear Studies   \\
The Graduate University for Advanced Studies \\
1-1, Ooho, Tsukuba, Ibaraki, 305-0801, Japan \\
E-mail: shunzo.kumano@kek.jp}

\address{$^\dagger$
Department of Physics, Kyoto University, 
Kyoto, 606-8502, Japan \\
E-mail: saito@nh.scphys.kyoto-u.ac.jp}

\maketitle

\abstracts{ 
We investigate the polarized gluon distribution $\Delta g(x)$
by a global analysis of current DIS data and the $\pi^0$ data
from RHIC-Spin experiments. The $\pi^0$ data provide a strong
constraint on $\Delta g(x)$, so that its uncertainty is reduced.
Furthermore, new DIS data of COMPASS and HERMES play an important
role in determining $\Delta g(x)$ at large $x$.}

\begin{picture}
(5,2)(-245,-410)
\put(0,-10){KEK-TH-1093}
\end{picture}
\vspace{-10mm}
\section{Introduction}
We have investigated the polarized parton distribution functions
(polarized PDFs) by a global analysis with deep inelastic scattering
(DIS) data.\cite{aac,aac06} The polarized valence-up and -down distributions,
$\Delta u_v(x)$ and $\Delta d_v(x)$, are determined well; however,
the antiquark and gluon distributions, $\Delta \bar{q}(x)$ and $\Delta g(x)$, 
have rather large uncertainties. It is therefore expected that $\Delta g(x)$
can be extracted from collider data in which gluon contributions dominate.
Fortunately, double spin asymmetry for the $\pi^0$ production is measured 
by the PHENIX collaboration at RHIC.\cite{phenix-pi-06} Because we are
interested in a possible constraint on $\Delta g(x)$ from the asymmetry data,
an attempt is made to determine $\Delta g(x)$ by an analysis including
the new data.

\section{Global analyses of the polarized PDFs and Results}
In this analysis, we chose the following functional form as a polarized PDF
at the initial $Q^2$ ($\equiv Q_0^2$):
\begin{equation}
   \Delta f(x,Q_0^2) 
   = [\delta x^{\nu}-\kappa (x^{\nu}-x^{\mu})] f(x,Q_0^2) \, ,
\label{eqn:df}
\end{equation}
where $\delta$, $\kappa$, $\nu$, and $\mu$ are free parameters, and
$f(x)$ is the unpolarized PDF. The positivity condition
$|\Delta f(x,Q_0^2))| \le f(x,Q_0^2)$ is imposed as a constraint
especially on the large-$x$ behavior of the polarized PDFs.
Moreover, the antiquark distributions are assumed to be flavor $SU(3)$
symmetric due to the lack of accuracy in present experimental data for
flavor separation of these distributions. We prepare four type distributions: 
$\Delta u_v(x)$, $\Delta d_v(x)$, $\Delta \bar{q}(x)$ and $\Delta g(x)$.
$Q^2$ dependence of the PDFs is taken into account by solving the DGLAP
equations numerically. The polarized PDFs, strictly speaking the free
parameters in Eq.(\ref{eqn:df}), are determined by a $\chi^2$ analysis
in the next-to-leading order (NLO). Uncertainties of these distributions
are estimated by the Hessian method. More details are found in
Ref.~\refcite{aac06} on the analysis method and the uncertainty
estimation.

We performed two analyses. The one is by using only the DIS data,
which include recent ones from COMPASS, HERMES, and JLab
experiments.\cite{compass-g1d,hermes-g1d,jlab-g1n} The other is by
using the $\pi^0$ asymmetry data of RUN05 at RHIC \cite{phenix-pi-06}
in addition to the DIS data. The total number of the experimental
data is 421, in which 413 and 8 are for the DIS and $\pi^0$ data,
respectively. The value of total $\chi^2 (/d.o.f.)$ is 358 (0.89)
for the analysis with only the DIS data, and it is 370 (0.90) for
the one with the $\pi^0$ and DIS data.

\begin{figure}[t]
	\centerline{\epsfxsize=45mm \epsfbox{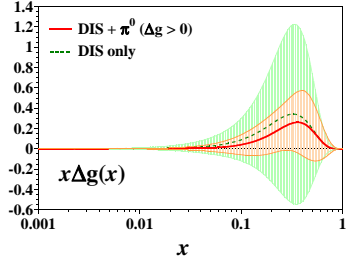}}
\vspace{-1mm}
\caption{Comparison of the polarized gluon distributions and their
uncertainties. The solid and dashed curves show $x\Delta g(x)$
from the DIS and $\pi^0$ asymmetry data and from only the DIS data,
respectively. The shaded areas are their uncertainties.
\label{xdgvsDIS}}
\vspace{-1mm}
\end{figure}

From these analyses, we obtain the first moments of $\Delta g(x)$:
\begin{eqnarray*}
	\Delta G(DIS+\pi^0) &=& 0.31 \pm 0.32 , \\
	\Delta G(DIS) &=& 0.47 \pm 1.08 . 
\end{eqnarray*}
The uncertainty is significantly reduced if the $\pi^0$ data are used
although the center value is slightly varied. Next, both polarized
gluon distributions are compared in Fig. \ref{xdgvsDIS}.
The $\Delta g(x)$ changes in the region $0.03<x<0.5$,
and its uncertainty is much reduced because of the $\pi^0$ data.
It indicates that the extraction of $\Delta g(x)$ from
the DIS data is difficult because the gluon contribution to $g_1(x)$
is rather small. It contributes indirectly through the $Q^2$ evolution and 
a higher-order correction. However, the $\pi^0$ production data are
useful to improve the precision of $\Delta g(x)$ determination.

In the analysis with the $\pi^0$ asymmetry data, there is, however,
a problem of $\Delta g(x)$ sign. The polarized cross section is roughly
proportional to square of $\Delta g(x)$ because the $gg$ scattering process
dominates in the low-$p_T$ region. Therefore, two types of solutions are
allowed: positive and negative $\Delta g(x)$. In practice, we perform
an analysis of a negative $\Delta g(x)$ input as an initial condition.
The value of the minimum $\chi^2$ for the $\pi^0$ data becomes 11.05 
which is almost the same as the one of the the positive $\Delta g(x)$
input. Therefore, the sign cannot be determined by the $\chi^2$ values.

\begin{figure}[b]
\vspace{-2mm}
	\centerline{\epsfxsize=65mm \epsfbox{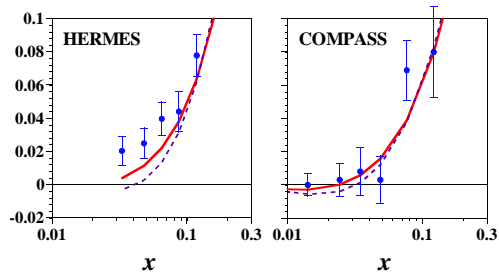}}
\vspace{-1mm}
\caption{Comparisons of the AAC fit with the asymmetry data $A_1^d(x,Q^2)$
of the HERMES and COMPASS.$^2$
\label{A1dHvsC}}
\end{figure}
In addition, we found an interesting fact that $\Delta g(x)$ becomes
positive at large $x$ in both cases. It is caused by the DIS data
of the HERMES and COMPASS experiments. Figure \ref{A1dHvsC} shows
comparison of the AAC fitting results with the asymmetry data of
deuteron target. Solid curves are full NLO calculation, and dotted
curves are obtained by eliminating the NLO gluon term from $g_1^d(x)$.
The differences between these curves indicate the gluon contribution
as a higher-order correction. For the COMPASS kinematics, the differences
are small because the $Q^2$ values are larger than those of the HERMES and
the NLO correction is smaller. However, the $Q^2$ values for the HERMES
data are a few GeV$^2$, so that the NLO contribution is rather large.
The differences are significant in the region $0.02<x<0.1$ where deviations 
between fit results and data exist. In order to explain the data,
the NLO gluon term should be positive.

\begin{figure}[t]
	\centerline{\epsfxsize=45mm \epsfbox{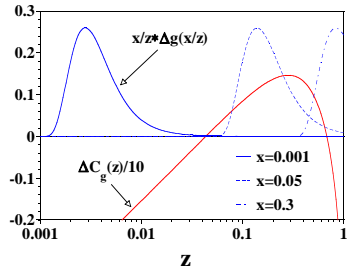}}
\vspace{-1mm}
\caption{$z$ dependence of the polarized gluon coefficient function
$\Delta C_g (z)$ and gluon distribution $\Delta g(x/z)$ in
the convolution integral for $g_1(x)$.$^2$
\label{Ceffivsxdg}}
\vspace{-1mm}
\end{figure}
For obtaining a positive gluon term in such an $x$ region, $\Delta g(x)$
must be positive at large $x$. The gluon term is given by the convolution
integral with the coefficient function $\Delta C_g (z)$:
$ \int^{1}_{x_{min}} dz/z \Delta C_g (z) \Delta g(x/z)$.
The behavior of these functions is shown in Fig. \ref{Ceffivsxdg}.
The coefficient function is positive in the region $0.02<z<0.7$.
To obtain the positive gluon term at $x=0.05$ where the deviation from
the HERES data exists, the gluon distribution must be positive
in the same $z$ region as shown by the dotted curve. The distribution
$\Delta g(0.05/z)$ in the region $0.05<z<0.1$ corresponds to $\Delta g(x)$
in the region $0.5<x<1$. Therefore, the gluon distribution should be
positive at large $x$ for fitting to the experimental data.

\vspace{-1.5mm}
\section{Summary}
\vspace{-0.5mm}
For determination of $\Delta g(x)$, we performed the global analyses
with present DIS and $\pi^0$ production data. Although the uncertainty
of the fist moment is significantly reduced by adding the $\pi^0$ data,
the sign problem of $\Delta g(x)$ appears. However, the DIS data of
COMPASS and HERMES experiments provide a constraint on the large-$x$
behavior of $\Delta g(x)$ through the NLO gluon correction term in $g_1(x)$.

\end{document}